\documentclass[twocolumn,showpacs,amsmath,amssymb]{revtex4}
\usepackage{graphicx}
\usepackage{dcolumn}
\usepackage{bm}

\begin{document}

\title{
Fault-tolerant quantum computation
in concatenation of verified cluster states
}

\author{Keisuke Fujii}
\author{Katsuji Yamamoto}
\affiliation{
Department of Nuclear Engineering, Kyoto University, Kyoto 606-8501, Japan}

\date{\today}

\begin{abstract}
A novel scheme is presented for fault-tolerant quantum computation
based on the cluster model.
Some relevant logical cluster states are constructed
in concatenation by post-selection through verification,
without necessity of recovery operation,
where a suitable code such as the Steane's 7-qubit code
is adopted for transversal operations.
This simple concatenated construction of verified cluster states
achieves a high noise threshold $ \sim 1 \% $,
and restrains the divergence of resources.
\end{abstract}


\maketitle

In order to implement reliable computation in physical systems,
the problem of noise should be overcome.
Then, fault-tolerant quantum computation
with error correction has been investigated
\cite{Shor95,CaldShor96,Shor96,Stea96,DiViShor96,Stea97,Stea98}.
In the usual quantum error correction (QEC), error syndromes are detected
on encoded qubits, and the errors are corrected according to them.
The noise thresholds for fault-tolerant computation are calculated
to be about $10^{-6} - 10^{-3}$
depending on the QEC protocols and noise models
\cite{Stea97,Stea98,Stea99,Stea03,Kitaev97,Preskill98,Knill98,
Gottesman97,Gottesman98,AB-O99}.
A main motivation for QEC comes from the fact that
in the circuit model the original qubits should be used
throughout computation even if errors occur on them.

On the other hand, more robust computation may be performed
in measurement-based quantum computers
\cite{GC99,ZLC00,OWC,Niel03,Tame07,Knill05a,Knill05b}.
Teleportation from old qubits to fresh ones
is made by measurements for gate operations,
and the original qubits are not retained.
An interesting computation model
with error-correcting teleportation is proposed
based on encoded Bell pair preparation and Bell measurement,
which provides high noise thresholds $ \sim 1 - 3 \% $
\cite{Knill05a,Knill05b}.
The cluster model or one-way computer \cite{OWC}
should also be considered for fault-tolerant computation.
A highly entangled state, called a cluster state, is prepared,
and gate operations are implemented
by measuring the qubits in the cluster
with feedforward for the post-selection of measurement bases.
This gate operation in the cluster model
may be viewed as the one-bit teleportation \cite{ZLC00}.
A promising scheme for linear optical quantum computation is proposed,
where deterministic gates are implemented
by means of the cluster model \cite{Niel04}.
Fault-tolerant computation is built up for this optical scheme
by using a clusterized version of the syndrome extraction for QEC
\cite{Stea97}.
The noise thresholds are estimated to be about $ 10^{-3} $ for photon loss
and $ 10^{-4} $ for depolarization \cite{Niel06}.
The threshold result is also argued
by simulating the QEC circuits with clusters
\cite{Rausen03,ND05,AL06}.
Some direct approaches are, on the other hand, considered
for the fault-tolerant one-way computation
\cite{FY06,Silva07}.

In this Letter, we present a novel scheme
of fault-tolerant quantum computation
by making a better use of the unique feature of the cluster model.
Specifically, the fault-tolerant computation
is implemented by {\it concatenated construction and verification
of logical cluster states
via one-way computation with post-selection}.
A number of cluster states are constructed
in parallel with error detection,
and the unsuccessful ones are discarded,
selecting clean cluster states.
The high-fidelity preparation of Bell state (or its cluster version)
is adopted for the error-correcting teleportation
\cite{Knill05a,Knill05b,Silva07}.
It is also considered that improved ancilla preparation
increases the noise threshold \cite{Reichardt04,Eastin07}.
In the present scheme, even gate operations as cluster states
are prepared and verified by post-selected computing
to reduce errors more efficiently.
That is, gate operations are {\it pre-selected},
or errors are corrected before the computation starts,
say {\it error pre-correction},
which is enabled by means of the cluster model
where the order of operations can be changed suitably
(see Ref. \cite{FY06} for an early idea).
This is quite distinct from the standard QEC,
where errors are corrected after noisy operations,
even via teleportation.

While high-fidelity state preparation is achieved by post-selection,
huge resources are generally required
due to the exponentially diminishing net success probability
according to the computation size,
which is a serious obstacle for scalability
\cite{Knill05a,Knill05b,FY06,Silva07}.
We here succeed to overcome this dilemma in post-selection
by presenting a systematic method of concatenation
to construct logical cluster states through verification.
As described in the following, the necessary post-selections
are minimized and localized,
which enables off-line gate operations prior to the computation,
as verified logical clusters.
This provides the scalable concatenation of post-selection
in the cluster model.
Then, a high noise threshold $ \sim 1 \% $ is achieved
by post-selection, while the resources usage is moderate,
being comparable with or even less than the circuit-based QEC schemes.
This concatenated cluster construction is implemented suitably
by adopting a class of stabilizer codes
of Calderbank-Shor-Steane, e.g., the Steane's 7-qubit code
\cite{CaldShor96,Stea96,Gottesman98}.
The logical measurements of Pauli operators
as well as the Clifford gates, $ H $, $ S $ and C-$ Z $,
are implemented transversally on such a quantum code.
The non-Clifford $ \pi / 8 $ gate is even operated
for universal computation by preparing a specific qubit
and making a transversal measurement
\cite{FY06,Silva07}.

(i) {\it Fundamental clusters:}
A set of gate operations in one-way computation may be decomposed
into some fundamental clusters.
This decomposition enables us to post-select the computation
without divergence of resources in concatenation.
The fundamental clusters are specifically taken as
$ | h^{(l)} \rangle $, $ | 0^{(l)} \rangle $, $ | +^{(l)} \rangle $
at the logical level $ l $ ($ l \geq 1 $),
which are composed of level-$ (l-1) $ qubits.
The code states $ | 0^{(l)} \rangle $ and $ | +^{(l)} \rangle $
are used as ancillas for encoding and syndrome detection.
The hexa-cluster $ | h^{(l)} \rangle $,
as a linear cluster of 6 qubits,
represents an elementary unit of gate operations.
These fundamental clusters are combined
by {\it bare} C-$ Z $ gates
(transversal concatenation of physical C-$ Z $ gates without verification)
to implement one-way computations
such as C-$ Z $ gates with syndrome detections
and the concatenated construction of logical clusters
through verification,
$ \{ | h^{(l)} \rangle , | 0^{(l)} \rangle , | +^{(l)} \rangle \}
\rightarrow
\{ | h^{(l+1)} \rangle , | 0^{(l+1)} \rangle , | +^{(l+1)} \rangle \} $.

\begin{figure}
\centering
\scalebox{.3}{\includegraphics*[0cm,0.5cm][30cm,10cm]{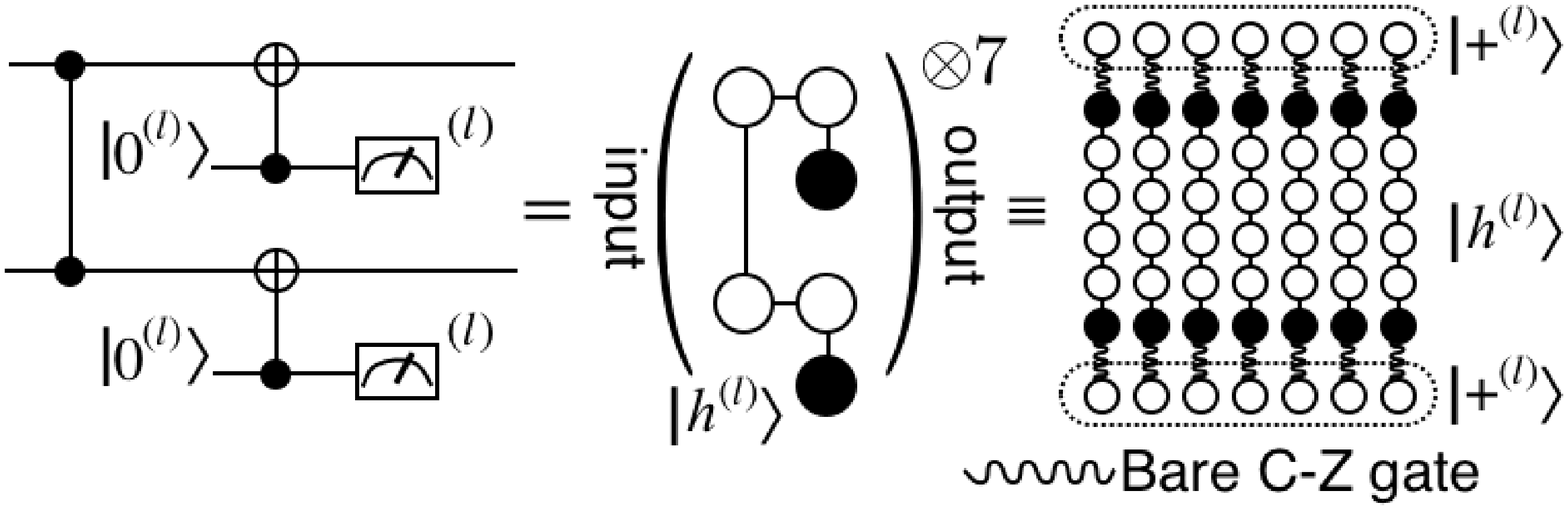}}
\caption{C-$ Z $ gate with single verification at the level $ l $.
Each $ | +^{(l)} \rangle $ (= $ H | 0^{(l)} \rangle $)
is encoded through an $ H $ rotation to the level-$ (l-1)
$ {\LARGE $ \bullet $} qubits in 7 $ | h^{(l)} \rangle $'s
with bare C-$ Z $ gates.}
\label{single}
\end{figure}

\begin{figure}
\centering
\scalebox{.275}{\includegraphics*[0cm,0.5cm][31.5cm,10cm]{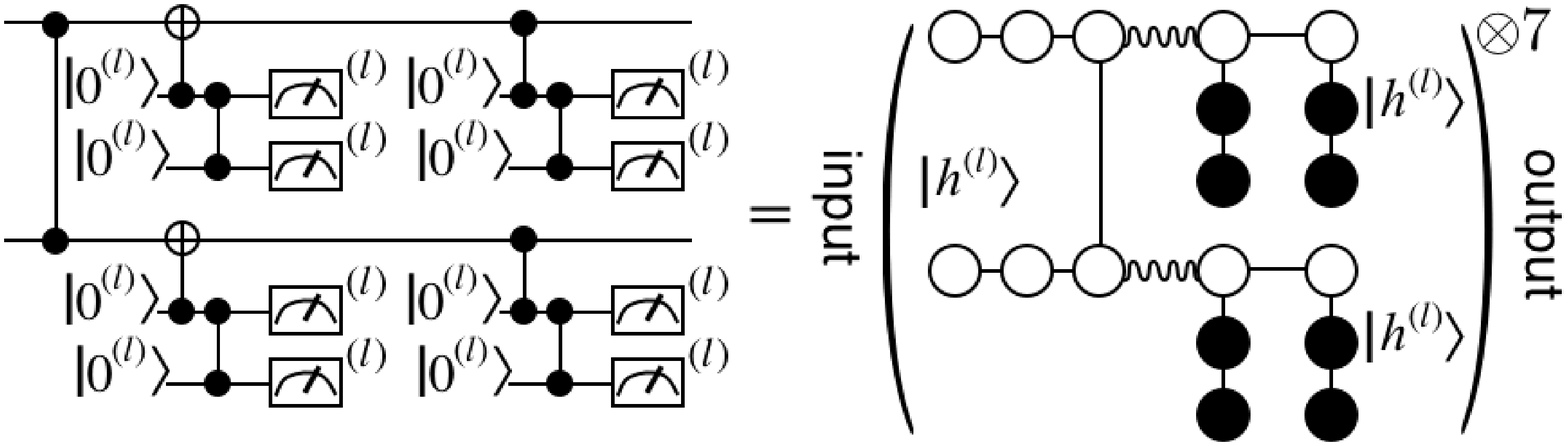}}
\caption{C-$ Z $ gate with double verification at the level $ l $.
Ancilla $ | +^{(l)} \rangle $'s should be encoded
to the {\LARGE $ \bullet $} qubits as Fig. \ref{single}.}
\label{double}
\end{figure}

\begin{figure}
\centering
\scalebox{.25}{\includegraphics*[0cm,0.75cm][27.5cm,15.25cm]{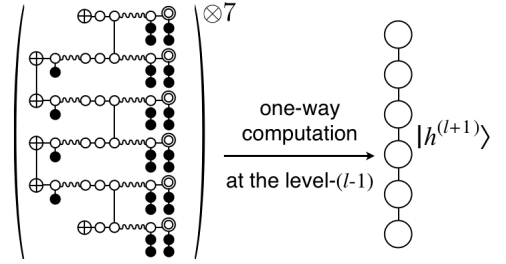}}
\caption{Concatenated construction of hexa-cluster.}
\label{hexa}
\end{figure}

(ii) {\it Verified C-$ Z $ gates:}
A C-$ Z $ gate with single verification at the level-$ l $
is implemented by combining $ 7 | h^{(l)} \rangle $'s
and $ 2 | +^{(l)} \rangle $'s,
as shown in the right cluster diagram of Fig. \ref{single}.
This combination of clusters is schematically
denoted by the symbol ``$ \otimes 7 $"
(henceforth used conveniently),
which does not simply imply the tensor product
but also includes the encoding of ancilla code blocks
(marked with {\LARGE $ \bullet $}).
The level-$ (l-1) $ {\LARGE $ \bullet $} qubit
in each $ | h^{(l)} \rangle $ is connected through an $ H $ rotation
to the corresponding level-$ (l-1) $ qubit
in an ancilla $ | +^{(l)} \rangle $
with a bare C-$ Z $ gate (wavy line);
$ | +^{(l)} \rangle = H | 0^{(l)} \rangle $
is teleported as $ | 0^{(l)} \rangle $.
The 2 input level-$ l $ code blocks are similarly encoded
via teleportation to the $ 2 \times 7 $ level-$ (l-1) $
{\small $ \bigcirc $} qubits
(see also $ \oplus $'s in Fig. \ref{hexa}).
As seen from the circuit equivalent in Fig. \ref{single},
the error syndromes of the 2 level-$ l $ qubits
through the C-$ Z $ gate are extracted for verification
\cite{Stea97,Stea98,Stea99}.
A C-$ Z $ gate with double verification is also implemented
by combining $ 7 \times 3 $ $ | h^{(l)} \rangle $'s
and $ 8 | +^{(l)} \rangle $'s in Fig. \ref{double}
(the ancillas are encoded to the {\LARGE $ \bullet $} qubits).
Here, the errors in the ancilla $ | +^{(l)} \rangle $'s
($ | 0^{(l)} \rangle $'s via teleportation)
are even detected for higher fidelity.
It should be remarked that at the beginning of concatenation
the verified level-1 C-$ Z $ gates may be implemented efficiently
by means of the circuit diagrams in Figs. \ref{single} and \ref{double},
without using $ | h^{(1)} \rangle $'s.
This is because $ | h^{(1)} \rangle $'s, as chains of physical qubits
without verification, are somewhat noisy.

(iii) {\it Concatenated cluster construction:}
The level-$ (l+1) $ hexa-cluster $ | h^{(l+1)} \rangle $
is constructed in Fig. \ref{hexa}
by combining the level-$ l $ clusters
$ | h^{(l)} \rangle $, $ | 0^{(l)} \rangle $, $ | +^{(l)} \rangle $
with bare C-$ Z $ gates.
Here, it is understood that
the 7 level-$ (l-1) $ $ \oplus $ qubits
in a transversal set of 7 $ | h^{(l)} \rangle $'s
are connected to a $ | 0^{(l)} \rangle $ (not shown explicitly)
with bare C-$ Z $ gates to encode
$ | +^{(l)} \rangle = H | 0^{(l)} \rangle $
via teleportation, as done similarly for the {\LARGE $ \bullet $} qubits
in Fig. \ref{single}.
The 6 $ | +^{(l)} \rangle $'s encoded
to the $ 6 \times 7 $ $ \oplus $ qubits
are entangled through 2 C-$ Z $ gates with single verification
(Fig. \ref{single})
and 3 C-$ Z $ gates with double verification (Fig. \ref{double})
to form the $ | h^{(l+1)} \rangle $ via one-way computation.
These C-$ Z $ gates are combined in such ways
that each qubit has at most one bare C-$ Z $ connection
(wavy line), and that the output qubits ({\large $ \circledcirc $})
as $ | h^{(l+1)} \rangle $ are doubly verified.
The level-$ (l-1) $ qubits, except {\large $ \circledcirc $}'s,
are measured to implement the computation.
In this transversal level-$ (l-1) $ computation,
the level-$ l $ syndromes can be extracted by the measurements
of the level-$ (l-1) $ {\LARGE $ \bullet $} qubits.
Then, if all the level-$ l $ syndromes are correct,
the $ 6 \times 7 $ level-$ (l-1) $ {\large $ \circledcirc $} qubits survive
as the verified $ | h^{(l+1)} \rangle $ passing the post-selection.
(Once an error syndrome is detected, the computation is abandoned,
among many parallel constructions.)
That is, the level-$ l $ gate operation to be implemented
with $ | h^{(l+1)} \rangle $ has been verified beforehand
by these level-$ l $ syndrome extractions.
As noted previously, the level-2 construction may be implemented
efficiently with the circuit diagrams
in Figs. \ref{single} and \ref{double}
for the verified level-1 C-$ Z $ gates.

The level-1 $ | 0^{(1)} \rangle $ and $ | +^{(1)} \rangle $ are prepared
through verification by the stabilizer measurement and syndrome extraction
with the usual method
\cite{Stea97,Stea98,Stea99}.
Then, the level-$ (l+1) $ $ | +^{(l+1)} \rangle $
and $ | 0^{(l+1)} \rangle $ are encoded
with the level-$ l $ fundamental clusters
via one-way computation in cluster diagrams,
such as Fig. \ref{hexa} for the $ | h^{(l+1)} \rangle $,
which include suitably the verified C-$ Z $ gates
(Figs. \ref{single} and \ref{double}).

(iv) {\it Universal computation:}
The fundamental clusters are constructed through verification
up to the highest logical level $ {\bar l} $
to achieve the fidelity required for a given computation size.
Then, the desired computation is implemented
by combining the highest-level hexa-clusters
$ | h^{({\bar l}+1)} \rangle $ with bare C-$ Z $ gates.
The preparation of $ | \pi / 8^{({\bar l})} \rangle
= \cos ( \pi / 8 ) | 0^{({\bar l})} \rangle
+ \sin ( \pi / 8 ) | 1^{({\bar l})} \rangle $
is also needed for universal computation
to operate the non-Clifford $ \pi / 8 $ gate
$ e^{-i ( \pi / 8 ) Z} $ by transversal measurement on the 7-qubit code
\cite{FY06,Silva07}.
The level-1 $ | {\pi / 8}^{(1)} \rangle $
is encoded by the usual method \cite{Knill98}.
Then, the upper-level $ | {\pi / 8}^{(l)} \rangle $
is encoded with the lower-level $ | {\pi / 8}^{(l-1)} \rangle $,
similarly to the other fundamental clusters.
The logical failure of $ | {5 \pi / 8}^{(l)}\rangle $
cannot be detected in encoding the $ | {\pi / 8}^{(l)} \rangle $,
because it has also the correct syndrome.
This small mixture of $ | {5 \pi/8}^{(l)} \rangle $
is hence not reduced by the concatenation,
though the constructed $ | {\pi / 8}^{(l)} \rangle $
is kept on the code space by the verification,
retaining the logical fidelity
as the $ | {\pi / 8}^{(1)} \rangle $.
This slightly noisy $ | {\pi / 8}^{({\bar l})} \rangle $
is even useful to obtain the desired high fidelity
$ | {\pi / 8}^{({\bar l})} \rangle $ at the highest level
by using the magic state distillation with Clifford operations
\cite{BK05}.

The errors on the qubits and Pauli frames should be considered properly
to estimate the measurement errors and noise threshold
in the one-way computation with post-selection
to prepare clean logical clusters.

(i) {\it Homogeneous errors on qubits:}
The level-2 fundamental clusters are first constructed,
and their constituent level-1 qubits are doubly verified.
Then, it is reasonably expected that the level-0 qubits
(as marked with {\large $ \circledcirc $} in Fig. \ref{hexa})
encoded in these verified level-1 qubits
contain independently and identically distributed (homogeneous)
depolarization errors in the leading order \cite{Eastin07}.
The homogeneous error probabilities
$ \epsilon_A $ ($ A = X, Y, Z $) of these level-0 qubits
are determined essentially by the error probabilities $ p_{AB} $
of the physical gates which are used transversally
for the level-1 double verification;
$ \epsilon_X = p_{XI} $, $ \epsilon_Y = p_{YI} $,
$ \epsilon_Z = 2 p_{ZI} $
from the circuit diagram in Fig. \ref{double}.
(The errors on the input qubits are almost eliminated
through the verification.)
In the level-$ (l+1) $ construction, as described in Fig. \ref{hexa},
{\it any operations are not implemented directly
on the output level-$ l $ qubits,
which are composed of the level-$ (l-1) $ {\large $ \circledcirc $} qubits,
but the entanglement by the verified C-$ Z $ gates is transferred
to prepare the verified level-$ (l+1) $ clusters
via teleportation (one-way computation) of the level-$ (l-1) $ qubits}.
Hence, these output level-$ l $ qubits inherit transversally
the homogeneous errors $ \epsilon_A $ of the constituent level-0 qubits
after the level-1 verification.
The prepared level-$ (l+1) $ clusters are further used
for the level-$ (l+2) $ construction,
and some pairs of level-$ l $ qubits in these clusters
are connected by bare C-$ Z $ gates.
Then, extra errors are added to the constituent level-0 qubits
through the bare C-$ Z $ gate
as $ \epsilon^\prime_X = \epsilon_X + p_{XB} $,
$ \epsilon^\prime_Y = \epsilon_Y + p_{YB} $,
$ \epsilon^\prime_Z = \epsilon_Z + \epsilon_X + \epsilon_Y + p_{ZB} $
(summed over $ B = I, X, Y, Z $).

(ii) {\it Errors in measurements and threshold:}
By the measurements of level-$ (l-1) $ qubits
to construct the level-$ (l+1) $ clusters, as in Fig. \ref{hexa},
the level-$ (l-1) $ Pauli frames of the neighboring qubits are updated.
The output level-$ l $ qubits to form the level-$ (l+1) $ clusters
are, however, doubly verified,
and hence the propagation of the preceding measurement errors
is prohibited by post-selection
as the Pauli frame errors of the constituent level-$ (l-1)
$ {\large $ \circledcirc $} qubits.
The fundamental clusters are therefore prepared
to be free from the Pauli frame errors (up to the higher orders)
through the concatenation.
In the absence of Pauli frame errors at the level-$ (l-1) $ and below,
the error probability $ p_q^{(l)} $ to measure solely a level-$ l $ qubit
contained in a verified level-$ (l+1) $ cluster
is reduced transversally to the level-0 $ p_q^{(0)} $
on the 7-qubit code with distance 3 as
\begin{equation}
p_q^{(l)} \simeq {}_7 {\rm C}_2 ( p_q^{(l-1)} )^2
\simeq ( {}_7 {\rm C}_2 p_q^{(0)} )^{2^l} / {}_7 {\rm C}_2 .
\label{eqn:pq}
\end{equation}
The level-$ l $ qubit is actually measured
during the upper level-$ (l+2) $ cluster construction.
Then, the measurement error of this qubit becomes 
some multiple of $ p_q^{(l)} $, including its level-$ l $ Pauli frame error
due to the propagation of the preceding measurement errors.

The level-0 qubits with the homogeneous errors $ \epsilon^\prime_A $
through bare C-$ Z $ connection ($ \epsilon_A < \epsilon^\prime_A $)
are measured in the $ X $ basis with the error probability
$ p_q^{(0)} = \epsilon^\prime_Z + \epsilon^\prime_Y + p_M $
($ p_M $ is the error probability of physical measurement).
Then, the noise threshold is given from Eq. (\ref{eqn:pq}) as
\begin{equation}
p_q^{(0)} = D p_e < 1/{}_7 {\rm C}_2
\rightarrow p_{\rm th} = ( {}_7 {\rm C}_2 D )^{-1} ,
\label{eqn:pth}
\end{equation}
where $ p_e $ represents the mean error probability
of physical operations ($ D \sim 1 $).
It is estimated as $ p_{\rm th} = 0.042 $ ($ D = 17/15 $)
typically with $ p_{AB} = (1/15) p_e $
for $ \epsilon^\prime_{A} $ and $ p_M = (4/15) p_e $ \cite{Knill05b}.
We have made a numerical simulation to confirm the above estimates
concerning the errors on the qubits and Pauli frames
in the concatenation.
The Pauli frame errors are really absent
in the successful logical clusters in the leading order,
as considered in Eq. (\ref{eqn:pq}).

The physical resources (qubits and gates) are calculated
by counting the numbers of hexa-clusters, ancilla qubits
and bare C-$ Z $ gates in the diagrams
such as Figs. \ref{single}, \ref{double}, \ref{hexa}.
(The details will be presented in a forthcoming paper.)
They are given as the recurrence relations
for the C-$ Z $ gates with single ($ S $) and double ($ D $)
verifications, and the fundamental clusters
$ | \alpha \rangle = | h \rangle, | 0 \rangle , | + \rangle $:
\begin{eqnarray}
R_S^{(l)} &=& 7 R_h^{(l)} + 2 ( R_+^{(l)} + R_b^{(l)} )
( l \geq 2 ) ,
\label{eqn:RS}
\\
R_D^{(l)} &=& 3 \times 7 R_h^{(l)} + 8 ( R_+^{(l)} + R_b^{(l)} )
+ 2 R_b^{(l)} ( l \geq 2 ) ,
\label{eqn:RD}
\\
R_\alpha^{(l+1)}
&=& \sum_{O = S,D,0,b} n^O_\alpha R_O^{(l)} / p_\alpha^{(l+1)}
( l \geq 1 ) ,
\label{eqn:Ralpha}
\end{eqnarray}
where $ R_b^{(l)} = 7^l $ for a bare C-$ Z $ gate,
$ ( n^S_\alpha , n^D_\alpha , n^0_\alpha , n^b_\alpha ) $
= $ ( 2, 3, 6, 4 )_h $, $ ( 6, 7, 11, 15 )_0 $, $ ( 5, 6, 10, 14 )_+ $,
and the success probabilities $ p_\alpha^{(l+1)} $
for the cluster verification are included.
The level-1 resources are given by
$ R_0^{(1)} = 69 / p_0^{(1)} $,
$ R_+^{(1)} = 72 / p_+^{(1)} $,
$ R_S^{(1)} = 3 \times 7 + 2 R_0^{(1)} $,
$ R_D^{(1)} = 9 \times 7 + 8 R_0^{(1)} $
[$ n^0_\alpha R_0^{(1)} \rightarrow n^0_\alpha R_+^{(1)} $
for $ l = 1 $ in Eq. (\ref{eqn:Ralpha})],
based on the circuit diagrams in Figs. \ref{single} and \ref{double}
with physical C-Not and C-$ Z $ gates.
Somewhat more resources are used
if the cluster computation is made even at the level-0,
by substituting C-Not $ \rightarrow $ C-$ HZH $.

\begin{figure}
\centering
\scalebox{.75}{\includegraphics*[0cm,1.75cm][15cm,6.5cm]{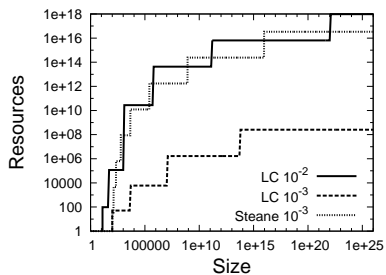}}
\caption{Resources for the present scheme of verified logical clusters
(LC) with $ p_e = 10^{-2} $ and $ 10^{-3} $,
which are compared with the Steane's QEC scheme with $ p_e = 10^{-3} $.
}
\label{resources}
\end{figure}

The success probabilities $ p_\alpha^{(l+1)} $
are evaluated by the numerical simulation,
which actually approach unity at the level-3 or higher
as the logical measurement error $ p_q^{(l)} $
is reduced rapidly below the threshold.
The resources are estimated in the above relations
with these $ p_\alpha^{(l+1)} $,
depending on the computation size $ N $
with the highest level $ {\bar l} \sim \log_2 ( \log_{10} N ) $
to achieve the accuracy $ 0.1/N $.
The results for $ R_0^{(l)} $ ($ > R_{h,+}^{(l)} $)
are shown in Fig. \ref{resources}
for the present scheme of verified logical clusters
(LC) with $ p_e = 10^{-2} $ and $ 10^{-3} $,
which are compared with the circuit-based Steane's QEC scheme
with $ p_e = 10^{-3} $ \cite{Stea03}.
Each step in these graphs means the up of logical level by one.
The present scheme really consumes much less resources
than the Steane's QEC scheme for $ p_e \leq 10^{-3} $.

We also find that compared with the $ C_{4}/C_{6} $ scheme
with post-selection (or with error-correction) \cite{Knill05a},
the present scheme provides a comparable threshold,
requiring much less (or comparable) resources.
Furthermore, the present scheme has a lot of room for improvement.
The Fibonacci scheme such as $ C_{4}/C_{6} $
based on the 4-qubit error-detecting code
may be applied to improve especially the resources.
The optimal decoding (adaptive concatenation)
\cite{Poulin06Fern08} is readily available
to boost the noise threshold up to $ 9 \% $
with reasonable resources.

The memory errors may be significant in this post-selection scheme
without recovery operation.
The qubits to form the clusters
are not touched directly (but via one-bit teleportation)
through the verified construction after the level-1 verification.
Then, the memory errors accumulate
until they are measured in the upper-level construction.
The memory errors are added as $ p_q^{(0)} + {\bar l}( n \tau_m p_e ) $,
where $ \tau_m p_e $ denotes the probability of memory error
with the effective waiting time $ \tau_m $ for one measurement,
and $ n $ is the number of waiting time steps
at each concatenation level (e.g., $ n = 12 $ for the hexa-cluster).
The noise threshold is hence determined as
$ p_{\rm th}
\sim [ {}_{7}{\rm C}_2 \{ 1 + \log_2 ( \log_{10} N ) n \tau_m \} ]^{-1} $,
depending on the computation size $ N $
with the highest level $ {\bar l} \sim \log_2 ( \log_{10} N ) $.
For example, $ p_{\rm th} \sim 1 \% $
for $ N \sim 10^{20} $ and $ \tau_m = 0.1 $ ($ n \sim 10 $),
which will be tolerable for practical computations.
In order to overcome essentially the memory error accumulation,
the fundamental clusters as two-colorable graph states
may be refreshed at each level
by using a purification protocol \cite{DAB03ADB05}.

This work was supported by International Communications Foundation (ICF).


\begin{thebibliography}{99}

\bibitem{Shor95}
P. W. Shor,
Phys. Rev. A {\bf 52}, R2493 (1995).

\bibitem{CaldShor96}
A. R. Calderbank and P. W. Shor,
Phys. Rev. A {\bf 54}, 1098 (1996).

\bibitem{Stea96}
A. M. Steane,
Phys. Rev. Lett. {\bf 77}, 793 (1996).

\bibitem{Shor96}
P. W. Shor,
{\it Proceedings of the 37th Annual Symposium
on Foundations of Computer Science}
(IEEE Computer Society Press, Los Alamitos, CA, 1996), p. 56.

\bibitem{DiViShor96}
D. P. DiVincenzo and P. W. Shor,
Phys. Rev. Lett. {\bf 77}, 3260 (1996).

\bibitem{Stea97}
A. M. Steane,
Phys. Rev. Lett. {\bf 78}, 2252 (1997).

\bibitem{Stea98}
A. M. Steane,
Fortschr. Phys. {\bf 46}, 443 (1998).

\bibitem{Stea99}
A. M. Steane,
Nature {\bf 399}, 124 (1999).

\bibitem{Stea03}
A. M. Steane,
Phys. Rev. A {\bf 68}, 042322 (2003).

\bibitem{Kitaev97}
A. Yu. Kitaev,
Russ. Math. Surv. {\bf 52}, 1191 (1997).

\bibitem{Preskill98}
J. Preskill,
Proc. R. Soc. London A {\bf 454}, 385 (1998).

\bibitem{Knill98}
E. Knill, R. Laflamme, and W. H. Zurek,
Proc. R. Soc. London A {\bf 454}, 365 (1998);
Science {\bf 279}, 342 (1998).

\bibitem{Gottesman97}
D. Gottesman, Ph.D. thesis, California Institute of Technology
(1997).

\bibitem{Gottesman98}
D. Gottesman,
Phys. Rev. A {\bf 57}, 127 (1998).

\bibitem{AB-O99}
D. Aharonov and M. Ben-Or,
{\it Proceedings of the 29th Annual ACM Symposium
on the Theory of Computation}
(ACM Press, New York, 1998), p. 176.

\bibitem{GC99}
D. Gottesman and I. L. Chuang,
Nature {\bf 402}, 390 (1999).

\bibitem{ZLC00}
X. Zhou, D. W. Leung, and I. L. Chuang,
Phys. Rev. A {\bf 62}, 052316 (2000).

\bibitem{OWC}
R. Raussendorf and H. J. Briegel,
Phys. Rev. Lett. {\bf 86}, 5188 (2001);
R. Raussendorf, D. E. Browne, and H. J. Briegel,
Phys. Rev. A {\bf 68}, 022312 (2003).

\bibitem{Niel03}
M. A. Nielsen,
Phys. Lett. A {\bf 308}, 96 (2003).

\bibitem{Tame07}
R. Prevedel, M. S. Tame, A. Stefanov, M. Paternostro,
M. S. Kim, and A. Zeilinger,
Phys. Rev. Lett. {\bf 99}, 250503 (2007).

\bibitem{Knill05a}
E. Knill,
Nature {\bf 434}, 39 (2005).

\bibitem{Knill05b}
E. Knill,
Phys. Rev. A {\bf 71}, 042322 (2005).

\bibitem{Niel04}
M. A. Nielsen,
Phys. Rev. Lett. {\bf 93}, 040503 (2004).

\bibitem{Niel06}
C. M. Dawson, H. L. Haselgrove, and M. A. Nielsen,
Phys. Rev. Lett. {\bf 96}, 020501 (2006);
Phys. Rev. A {\bf 73}, 052306 (2006).

\bibitem{Rausen03}
R. Raussendorf,
Ph.D. thesis, Ludwig-Maximillians Universit{\"a}t
M{\"u}nchen (2003).

\bibitem{ND05}
M. A. Nielsen and C. M. Dawson,
Phys. Rev. A {\bf 71}, 042323 (2005).

\bibitem{AL06}
P. Aliferis and D. W. Leung,
Phys. Rev. A {\bf 73}, 032308 (2006).

\bibitem{FY06}
K. Fujii and K. Yamamoto,
quant-ph/0611160 (2006).

\bibitem{Silva07}
M. Silva, V. Danos, E. Kashefi, and H. Ollivier,
New J. Phys. {\bf 9}, 192 (2007).

\bibitem{Reichardt04}
B. W. Reichardt, quant-ph/0406025 (2004).

\bibitem{Eastin07}
B. Eastin,
Phys. Rev. A {\bf 75}, 022301 (2007).

\bibitem{BK05}
S. Bravyi and A. Kitaev,
Phys. Rev. A {\bf 71}, 022316 (2005).

\bibitem{Poulin06Fern08}
D. Poulin,
Phys. Rev. A {\bf 74}, 052333 (2006);
J. Fern,
Phys. Rev. A {\bf 77}, 010301(R) (2008).

\bibitem{DAB03ADB05}
W. D{\"u}r, H. Aschauer, and H. J. Briegel,
Phys. Rev. Lett. {\bf 91}, 107903 (2003);
H. Aschauer, W. D{\"u}r, and H. J. Briegel,
Phys. Rev. A {\bf 71}, 012319 (2005).

\end{thebibliography}
\end{document}